\documentclass[preprint,12pt]{elsarticle}

\usepackage{amssymb}
\usepackage{amsmath}

\usepackage{algorithmic}
\usepackage{graphicx}
\usepackage{textcomp}

\usepackage{multirow}
\usepackage{comment}
\usepackage{amsfonts}
\usepackage{booktabs}
\usepackage{siunitx}
\usepackage{tabularx}

\journal{}

\begin{document}

\begin{frontmatter}



\title{Geodesic-based Predictive Shape Modeling of the Right Ventricle in Patients with Hypoplastic Left Heart Syndrome}



\author[1]{Ye Han}
\author[1]{James Fishbaugh}
\author[1]{Jared Vicory}
\author[2]{Silvani Amin}
\author[2]{Matthew Daemer}
\author[2]{Hannah E. Dewey}
\author[2]{Yan Wang}
\author[2]{Analise M. Sulentic}
\author[2]{Alana Cianciulli}
\author[3]{Andras Lasso}
\author[2]{Matthew A. Jolley}
\author[1]{Beatriz Paniagua}

\affiliation[1]{organization={Kitware},
            city={Clifton Park},
            postcode={12065}, 
            state={NY},
            country={USA}}
\affiliation[2]{organization={Children's Hospital of Philadelphia},
            city={Philadelphia},
            postcode={19104}, 
            state={PA},
            country={USA}}

\affiliation[3]{organization={Queen's University},
            city={Kingston},
            postcode={K7L 3N6}, 
            state={ON},
            country={Canada}}

\begin{abstract}
Hypoplastic left heart syndrome (HLHS) is characterized by severe underdevelopment of left ventricle requiring staged surgical reconstruction (stages) to allow the right  ventricle (RV) alone to support the circulation.  In this setting changes in RV size and shape over time reflect adaptations to single-ventricle physiology, dysfunction of the associated tricuspid valve (TV), and are associated with circulatory failure. As such, an accurate prediction of the RV shape of a patient would inform understanding of both RV and TV failure, as well as clinical prognosis and associated decision making. We present a geodesic-based predictive shape modeling framework applied a cohort of RVs obtained from 15 HLHS patients at three individual time points. Reasonable predictions on stage 1 RV shapes can generated using pre-stage 1 RV shapes and two predictors from prior clinical and demographic measures. Our results demonstrate the future potential for a data-driven method to predict how the morphology of the RV of an individual patient will change in size and shape over time. Future studies will seek to expand the training sample size and integrate more comprehensive demographic and morphological data into the proposed predictive model.
\end{abstract}




\end{frontmatter}






\section{Introduction}
\label{sec:introduction}
Hypoplastic left heart syndrome (HLHS) is a congenital defect where the left side structures of the heart are severely underdeveloped which affects 1 out of every 3,841 babies born \cite{dattahlhs2023}. HLHS is fatal without early surgical intervention, which takes the form of a series of open heart surgical procedures to allow the single right ventricle and associated tricuspid valve to support the systemic circulation \cite{rychik2019fontan}. It has been shown that tricuspid regurgitation (TR) is associated with RV dysfunction and is a significant contributor to mortality \cite{kutty2014tricuspid,ghanayem2012interstage,tabbutt2012risk,ugaki2013tricuspid}. Furthermore, because the tricuspid valve is integrated into the RV both at the outer rim of the valve (the annulus) and suspended in the RV by papillary muscles and chords, the two structures are highly interdependent. TV failure can result in adaptive ventricular dilation, and ventricular dilation can result in dilation of the valve annulus resulting in TV regurgitation \cite{spinner2011ftrcharacterization, colen2018tvadaptation}. One question of interest is the prediction of RV shape at different stages of surgical intervention as a co-indicator of both ventricular and valvular performance. Such a predictive model would provide an association between RV morphology and function, and potentially inform clinical prognosis and associated decision making.

Statistical shape models have shown great potential as biomarkers of pathology, growth or treatment in medical image analysis\cite{elhabian2022ssm}. The morphology of anatomical structures is a rich, high dimensional measure which can facilitate quantification of many properties about anatomy such as thickness, curvature, and localized variability which global measures such as volume can not capture.

In this study, we propose a geodesic-based shape analysis framework for subject-specific models of shape trajectories that are used for predicting the RV shape at future stages given an initial RV shape. Individual children are followed over time during treatment, which will be referred to as pre-stage 1 and stage 1 (representing surgical interventions\cite{feinstein2012hlhsstages}). We then present the theoretical foundation for the predictive model built around our shape trajectories model \cite{Han2023Hierarchical}, and the application of the proposed framework to the longitudinal cohort of children with HLHS, with combination of RV shape information, clinical scores such as TR severity and demographics representing growth such as change in body surface area (BSA).

\section{Methods}
\subsection{Data}
In this study, the end-systole RV shapes of pre-stage 1 and stage 1 in a cohort of 15 HLHS patients are acquired from 3D echocardiogram-based speckle tracking, which were then transferred into the TOMTEC version 4.6 (Tomtec Imaging Systems, Unterschleißheim, Germany) for creating the corresponding point based 3D models of the RV chamber. Demographic information and clinical scores during acquisition were recorded including HLHS stages, ages, body surface area (BSA), TR severity grade, etc.

This research study was conducted retrospectively using human subject data with approval from the Children's Hospital of Philadelphia Review Board.

\subsection{Geodesic Shape Analysis Background}
For completeness, we include a brief description of the core methodology which serves as the foundation for designing and building all the shape models contained in this paper. The methodology is based on fundamental aspects of Riemannian geometry, which an interested reader can learn more from the acclaimed work of do Carmo~\cite{do1976differential}. 

\subsubsection{Shape Space}
In this work, shape space is defined as the space where translation, rotation, and scaling has been removed via Procrustes alignment. This transforms a shape representation as $(x,y,z)$ coordinates in Euclidean space to a point on a high-dimensional hyper-sphere which we denote as a Riemannian manifold $M$.

\subsubsection{Geodesic Operations}
Due to the curvature of the manifold $M$, we need to define a few components which allow us to compute Euclidean-like operations which properly account for such curvature. These are based on following shortest paths along the manifold $M$ which are traversed at constant speed. Such shortest-path curves are called geodesics, which extend the concept of straight lines to spaces with curvature.

Intuitively, an \textbf{exponential map} is a function which takes as input a point on the manifold and a vector (direction and magnitude) and returns a new point on the manifold. This allows one to follow a geodesic defined by a starting location and direction and returns the point at the end of the shortest path. More formally, we define
$$Exp(p,v) = q$$
as the exponential map which returns point $q\in M$ by following the geodesic starting from $p \in M$ defined by tangent vector $v$.

The \textbf{log map} is the inverse of the exponential map. This function takes as input two points on the manifold and returns the tangent which connects the two points with a geodesic (shortest-path) curve. It is defined as
 $$Log(p,q) = v$$
where $p,q \in M$ and $v$ is the tangent vector which defines a geodesic from $p$ to $q$.

Now, there is a very natural way to define a \textbf{scalar distance} between two shapes in shape space $M$. We define distance as
$$d(p,q) = ||Log(p,q)||$$
which is the $L^2$ norm (magnitude) of the log map between $p,q\in M$.

The final component to define is \textbf{parallel transport}. This function takes a vector at a certain point and returns the resulting vector after following a geodesic curve. In Euclidean space there is no need for a parallel transport operation, since moving a vector from one point to another does not change its angle. However, a vector will rotate when following a path in a space with curvature. There we define parallel transport
$$\psi_{p\rightarrow q}(v)$$
as the operation which transports a tangent vector $v$ from $p$ to $q$.

\subsection{Predictive Shape Modeling}
\begin{figure}[!ht]
    \centering
    \includegraphics[width=1.0\textwidth]{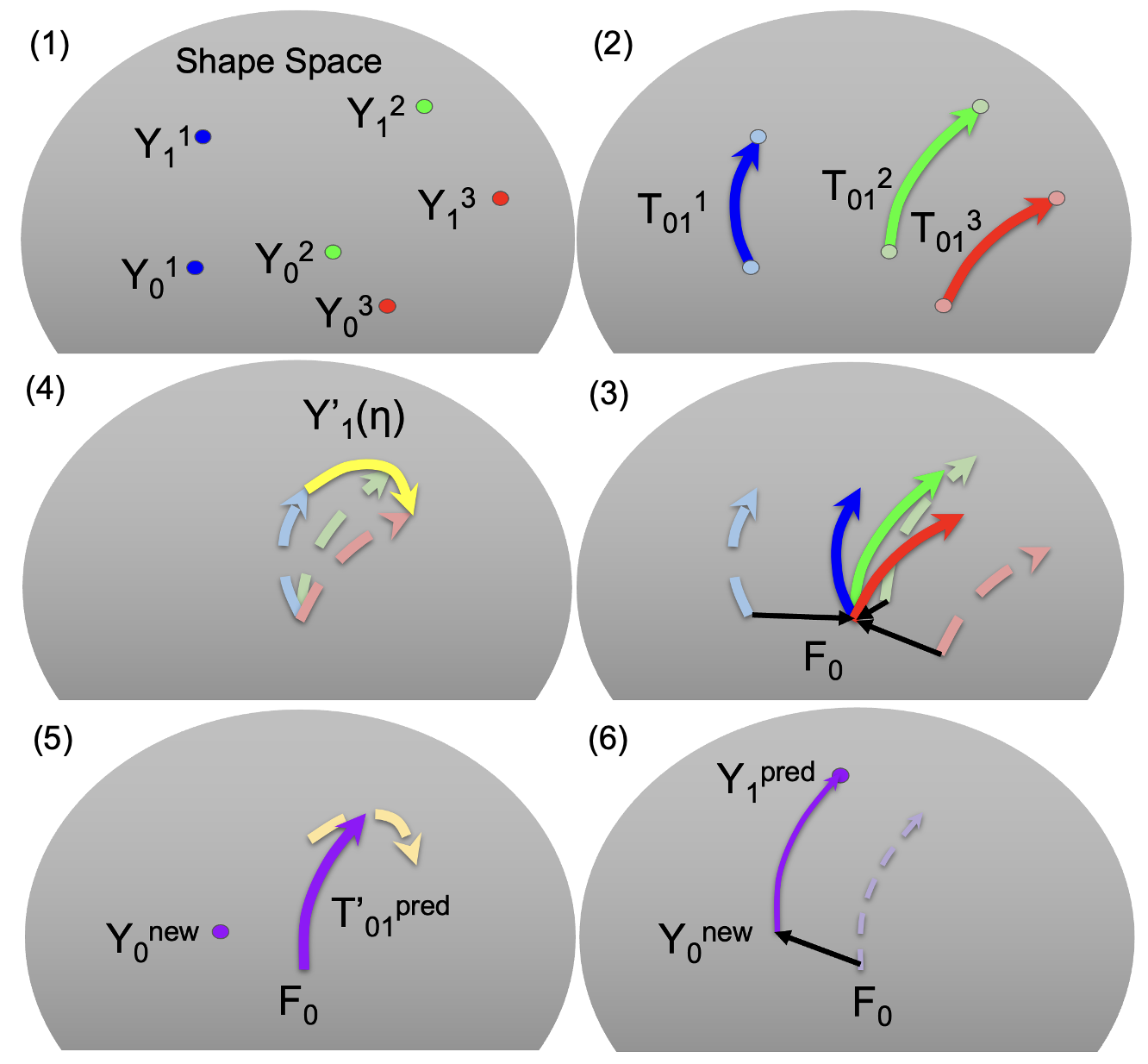}
    \caption{Pipeline of the predictive shape modeling framework based on geodesic models in a Riemannian shape space. (1) Longitudinal shapes for different subjects represented in shape space. (2) Geodesic models for individual subjects represent trajectory. (3) A common coordinate system is established by parallel transport. (4) A geodesic polynomial model is fit to the parallel translated parameters. (5) New subject can be predicted based on demographics. (6) Prediction is parallel transported back to original space of the subject.}
    \label{fig:pipeline}
\end{figure}

The overall pipeline of our predictive shape modeling framework is illustrated in Fig.\ref{fig:pipeline} which can be summarized as follows. (1) The shapes of interest are first preprocessed to establish spatial and temporal correspondence and transformed to shape space via partial Procrustes alignment. (2) A geodesic model is then fit to each subject to represent its trajectory of shape change over time. (3) The longitudinal models are aligned via parallel translation to the Fréchet mean $F_0$ of time $t_0$ shapes, to allow for shape modeling under the same coordinate frame in shape space. (4) A geodesic polynomial model is fit to the parallel translated longitudinal model parameters based on stage demographics. (5) When a new subject is presented, we predict new subject's longitudinal model parameters at $F_0$ based on its demographics. (6) We parallel translate predicted longitudinal model of the new subject to its time $t_0$ shape to finally get its predicted shape at stage of interest. More technical details of the individual steps are described in the subsections below.

\subsubsection{Shape Preprocessing}
In order to perform geodesic-based shape modeling, correspondence need to be established between pre-stage 1 and stage 1 shapes both within individual subjects and across the cohort population. We adopt the shape deformation based method \cite{Fishbaugh2018Correspondence} as it has shown the ability to handle shapes with thin and complex local geometry as well as being robust to topological changes. The method works by first defining a prototype shape configuration, which serves as the shared shape parameterization in sampling and topology for the population. Correspondence is then estimated by nonlinear shape registration between the prototype shape and each shape in the population. The RV shapes in correspondence are further transformed to a shape space where translation, rotation, and scaling have been removed via partial Procrustes alignment. This transforms a shape represented as $(x,y,z)$ coordinates in Euclidean space to a point on a high-dimensional hyper-sphere which we denote as a Riemannian manifold $M$ for geodesic-based shape modeling.

\subsubsection{Subject-wise Longitudinal Model}
We fit each subject-wise longitudinal trajectory with a unique geodesic model. The linear geodesic model is in the form of
\begin{equation}
Y_1^i = Exp(Y_0^i, T_{01}^i),
\end{equation}
where $Y_0^i$ and $Y_1^i$ are the time $t_0$ and time $t_1$ shapes of subject $i$ in shape space, and $T_{01}^i$ is the corresponding tangent vector at $Y_0^i$ parameterizing the geodesic change between the two shapes. $T_{01}^i$ can be obtained through our previously defined log map operation
\begin{equation}
T_{01}^i = Log(Y_0^i, Y_1^i).
\end{equation}
The goal here is to obtain the changes between stages (delta shapes) in shape space in the form of tangent vectors  (one per subject), which will serve as the objectives of the predictive model. If more intermediate time points (stages) are available, higher order polynomials can be used for subject-wise longitudinal shape modeling for more flexible and accurate shape matching.

\subsubsection{Parallel Translation to Fréchet Mean}
We parallel translate the subject-specific  tangent vectors to the Fréchet mean of time $t_0$ shapes $F_0$ to align them at the same local coordinate system. This procedure is necessary due to the curvature of the shape space, where a vector translated from one point to another also undergoes rotation. The translated tangent vectors are
$$\tilde{T}_{01}^i = \psi_{Y_0^i \rightarrow F_0}(T_{01}^i),$$
where $\psi_{Y_0^i \rightarrow F_0}$ is the parallel translation operator that translates vectors $T_{01}^i$ at $Y_0^i$ to their counterparts $\tilde{T}_{01}^i$ at $F_0$. Since the tangent vectors representing shape changes are defined at their respective time 0 points and thus not comparable, we need to transform them to the same location to allow for regression analysis. In our previous hierarchical longitudinal shape modeling work \cite{Han2023Hierarchical}, tangent vectors are translated to the same anchor point by following a two-step approach based on population level geodesic model and their covariate values. However, due to the heterogeneity of the possible time $t_0$ shapes, data points may be scattered in shape space. And it would not make sense to translate the tangent vector along specific covariate paths. Hence, we simply choose the Fréchet mean as the anchor point to translate tangent vectors to. Furthermore, the Fréchet mean is somehow the natural choice since it is the location which minimizes the sum of square distance to all shape observations, i.e. the middle.

\subsubsection{Geodesic Polynomial Model on Aligned Shapes}
Leveraging the transported tangent vectors, we generate aligned shape points by the exponential map
$$\tilde{Y}_1^i = Exp(F_0, \tilde{T}_{01}^i),$$
where $\tilde{Y}_1^i$ are the aligned shapes to be regressed on. This can be thought of as “shooting” the Fréchet mean along the parallel translated tangent vectors. The translated tangent vectors $\tilde{T}_{01}^i$ lie on the Euclidean hyper-tangent space at $F_0$ which does not allow for geodesic analysis in shape space. Hence, we generate aligned shape points using the above geodesic operation. We then perform geodesic polynomial regression on the generated shape points w.r.t. single/multiple covariates as described in the anchor point model in \cite{Han2023Hierarchical}. The final univariate geodesic polynomial model is in the form of
$$\tilde{y}_1^i = Exp(\hat{a_1}, \sum_{p=1}^n\hat{b_{1p}}{(c^i)}^p),$$
where $\tilde{y}_1^i$ is the prediction on $\tilde{Y}_1^i$, $\hat{a_1}$ and $\hat{b_{1p}}$ are regressed model parameters (location in shape space and tangent vectors), $n$ is the order of the polynomial, and $c^i$ is the regression variable. Interested readers may refer to \cite{Han2023Hierarchical} for the more complex multivariate versions of the geodesic polynomial model.

\subsubsection{Prediction and Error Quantification}
 Given a new subject $j$'s shape $Y_0^j$ at time 0 and its demographics $c^j$, the predicted shape $\tilde{y}_1^j$ at $F_0$ can be computed using the geodesic polynomial model above and its final prediction at time 1 can be obtained by
$$y_1^j = Exp(Y_0^j, \phi_{F_0 \rightarrow Y_0^j}(Log(F_0, \tilde{y}_1^j))).$$
Note that parallel translating the corresponding shapes from $F_0$ to $Y_0^i$ will not change their distance in shape space, thus the calculation of per-subject prediction error can be simplified as the geodesic distance between $\tilde{y}_1^i$ and $\tilde{Y}_1^i$ in the form of 
$${\epsilon}^i = d(\tilde{y}_1^i, \tilde{Y}_1^i). $$

\begin{table}[!b]
\centering
\caption{Prediction error in shape space based on various predictors: pre-stage 1 BSA (BSA\textsuperscript{0}), pre-stage 1 TR severity grade (TR\textsuperscript{0}), BSA difference ($\Delta$BSA), and the combination of TR severity and BSA difference. Baseline error is the surface distance between pre-stage 1 and stage 1.}\label{tab:prediction_errors}
\begin{tabular}{c|c|c|c|c|c} \toprule
\textbf{Case ID} & \textbf{Baseline} & \textbf{BSA\textsuperscript{0}} & \textbf{TR\textsuperscript{0}} & \textbf{$\Delta$BSA} & \textbf{TR\textsuperscript{0}/$\Delta$BSA} \\ \midrule
082  & 0.137 & 0.053 & 0.108 & 0.088 & 0.069 \\
169  & 0.192 & 0.130 & 0.128 & 0.147 & 0.110 \\
565  & 0.113 & 0.114 & 0.094 & 0.100 & 0.066 \\
627  & 0.225 & 0.177 & 0.206 & 0.201 & 0.205 \\ \midrule
630  & 0.102 & 0.108 & 0.074 & 0.083 & 0.069 \\
633  & 0.089 & 0.109 & 9.2x10$^{-5}$ & 0.074 & 1.4x10$^{-5}$ \\
641  & 0.219 & 0.278 & 0.228 & 0.227 & 0.224 \\
757  & 0.064 & 0.099 & 0.081 & 0.063 & 0.077 \\ \midrule
759  & 0.084 & 0.107 & 0.062 & 0.097 & 0.072 \\
766  & 0.096 & 0.085 & 0.076 & 0.059 & 0.063 \\
767  & 0.107 & 0.141 & 0.090 & 0.105 & 0.073 \\
837  & 0.162 & 0.097 & 0.110 & 0.121 & 0.107 \\ \midrule
1003 & 0.104 & 0.097 & 0.118 & 0.104 & 0.126 \\
1004 & 0.101 & 0.085 & 0.091 & 0.072 & 0.087 \\
1022 & 0.151 & 0.096 & 0.092 & 0.113 & 0.083 \\ \midrule
\textbf{mean} & \textbf{0.130} & \textbf{0.118} & \textbf{0.103} & \textbf{0.110} & \textbf{0.095} \\
\bottomrule
\end{tabular}
\end{table}
 
\section{Results and Discussion}
In this study, we use the RV shapes of pre-stage 1 and stage 1 in a cohort of 15 HLHS patients to demonstrate our predictive shape modeling framework. We experiment with using different combinations of pre-stage 1 RV shapes and demographic variables as predictors, including pre-stage 1 BSA, pre-stage 1 TR severity grade, BSA difference between pre-stage 1 and stage 1, and the combination of TR grade and BSA difference. All polynomial models are quadratic. The resulting prediction errors for various predictors are shown in respective columns in Table \ref{tab:prediction_errors} which are measured using the geodesic distance with unit “1”. Since the magnitude error and reconstruction measurements are normalized shape space metrics, it can be tough to understand what they mean. For this, we also include a baseline for comparison, which is the surface distance between the pre-stage 1 and stage 1 RV. This can be thought of a naive prediction that RV shape does not change over time. 

As can be seen from Table \ref{tab:prediction_errors}, using TR grades at pre-stage 1 yields overall smaller reconstruction error than using BSA at pre-stage 1 or BSA differences between the two stages. Combining TR grades and BSA differences to create a multivariate model further improves the overall prediction. For the majority of subjects, the prediction accuracy is considerably better than the baseline model.

\begin{figure}[!ht]
    \centering
    \includegraphics[width=1.0\textwidth]{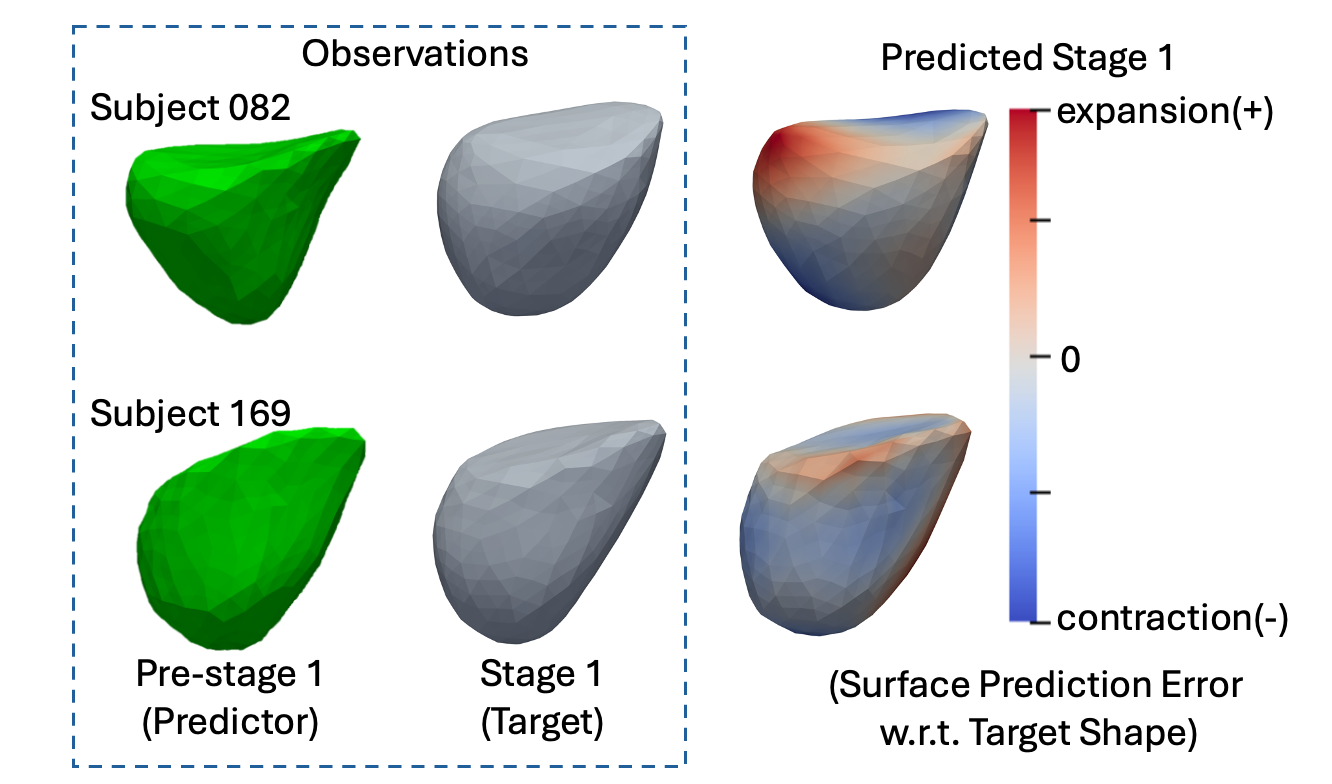}
    \caption{Predictions on two subjects' RV shapes at stage 1. The colormap represents surface prediction error in shape space. Our geodesic based shape prediction method is able to generate reasonably accurate and anatomically meaningful predictions on stage 1 RV shapes, with merely two predictors (BSA and TR grade). }
    \label{fig:visual_comparison}
\end{figure}

To visualize how the scale of reconstruction errors relate to the actual difference between shapes, visual comparisons between pre-stage 1, stage 1 and BSA-TR predicted shapes are created and shown in  Fig.\ref{fig:visual_comparison} for 2 example subjects. For most cases, our predictive model is capable of bringing pre-stage 1 shapes towards stage 1 shapes using only two predictors (BSA difference and TR grade). We believe we will be able to get further improved prediction results by incorporating more measured metrics in the demographics such as billow/tenting volume, annulus shape and/or coaptation surface characteristics.

\begin{figure}[!ht]
    \centering
    \includegraphics[width=0.85\textwidth]{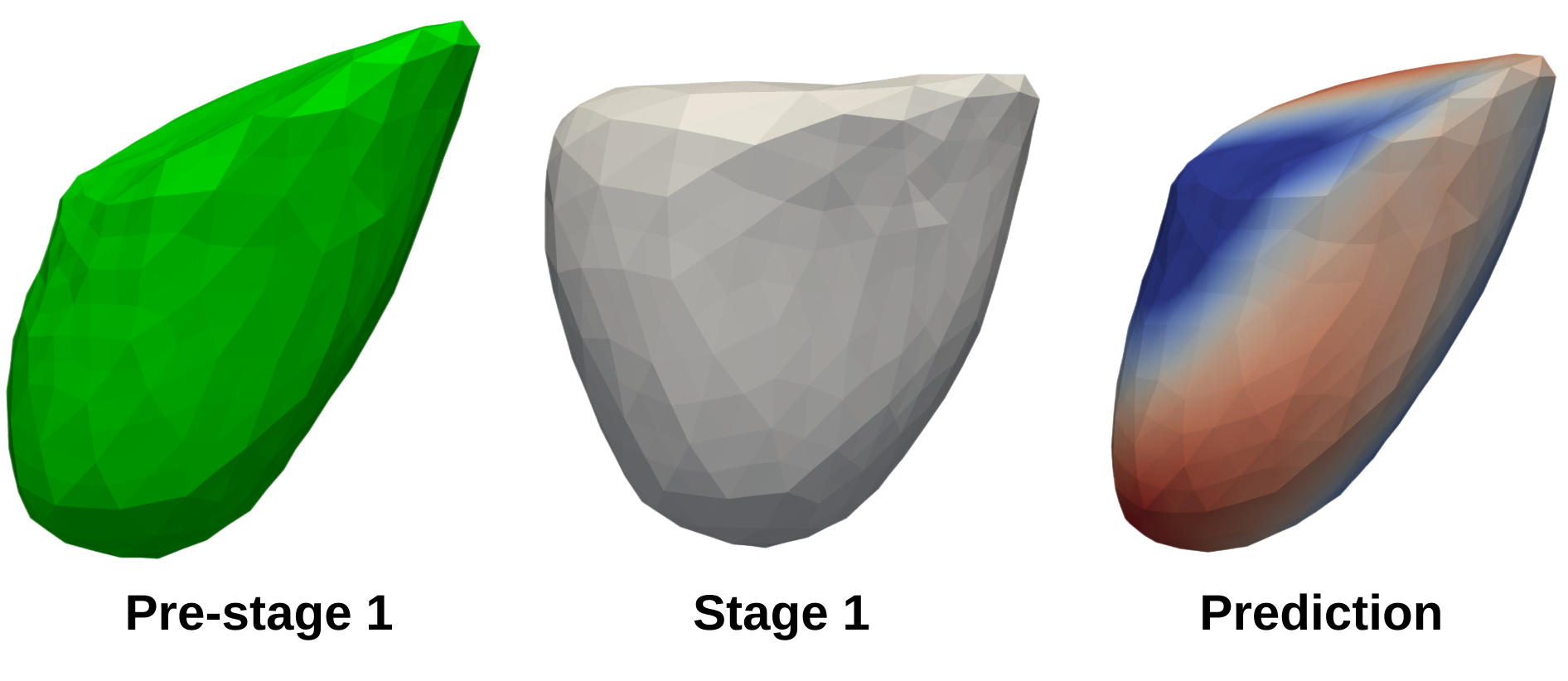}
    \caption{A case for which our predictive model did not perform well. It shows a significantly higher aspect ratio in its pre-stage 1 shape than those from the other subjects.}
    \label{fig:worst_case}
\end{figure}

For illustrative purpose, we also showcase a case (Fig.\ref{fig:worst_case}) for which the model did not perform well. We further look into the case and observed that its pre-stage 1 shape has a significantly higher aspect ratio than the other RV shapes in the dataset, which brings the necessity of incorporating more shape characteristics into our prediction model to account for the inherent pre-stage 1 shape variability in future studies.

\section{Future Work}
In this study, we preliminarily tested our predictive shape modeling framework on an initial cohort of RV shapes from HLHS patients. Based on the study results, future efforts will be made to justify and/or further improve the effectiveness of our modeling framework, which will include (1) incorporating a larger and more diverse patient cohort to validate the robustness and generalization of the model, (2) integrating additional clinical metrics, such as valve coaptation characteristics or ventricular wall stress into the model, and (3) add comparisons with other shape modeling or machine learning approaches that could highlight the superiority and unique advantages of the proposed framework.

\section{Conclusions}
We have demonstrated the feasibility of predicting RV shapes in patients with HLHS  using their previous-stage shapes and  limited demographic variables in a small cohort. We demonstrate that a multivariate model combining several clinical and demographic measures resulted in the highest predictive accuracy.  With further development application to a large cohort a model could be developed which allows the metrics from an individual patient to be placed within the context of the population.  This in turn will will inform accurate predictions to better understand clinical prognosis and support relevant procedural decisions.  However, further work will be needed demonstrate clinical impact on patient morbidity or mortality.

\section{Acknowledgments}
This research work was supported by the National Institute of Biomedical Imaging and Bioengineering, National Institute of Health, under award numbers R01HL153166 and R01EB021391.

\bibliographystyle{elsarticle-num}
\bibliography{bibliography.bib}

\end{document}